\documentclass{aastex62}

\usepackage{multirow}
\usepackage{fp}
\usepackage{url}
\usepackage{textcomp}
\usepackage{ulem}

\newcommand{\ergs}{erg\,s$^{-1}$}
\newcommand{\kms}{km\,s$^{-1}$}

\newcommand{\OBSTIME}{1156} 
\newcommand{\FLARENUMBER}{72} 
\newcommand{\FLARERATIO}{7.2} 
\newcommand{\ALPHALINEAR}{1.81\pm0.03} 
\newcommand{\ALPHAME}{1.52} 
\newcommand{\FLAREAREA}{$\approx 15-30\%$} 

\usepackage{color}

\graphicspath{{./}{figures/}}

\received{}
\revised{}
\accepted{}
\submitjournal{ApJ}

\shorttitle{Flaring activity of Proxima Centauri from TESS observations}
\shortauthors{Vida et al.}

\begin{document}

\title{Flaring activity of Proxima Centauri from TESS observations: quasi-periodic oscillations during flare decay and inferences on the habitability of Proxima b}

\correspondingauthor{Kriszti\'an Vida}
\email{vidakris@konkoly.hu}

\author[0000-0002-6471-8607]{Kriszti\'an Vida}
\affiliation{Konkoly Observatory, MTA CSFK, H-1121 Budapest, Konkoly Thege M. \'ut 15-17, Hungary}

\author[0000-0003-3669-7201]{Katalin Ol\'ah}
\affiliation{Konkoly Observatory, MTA CSFK, H-1121 Budapest, Konkoly Thege M. \'ut 15-17, Hungary}

\author[0000-0001-5160-307X]{Zsolt K\H{o}v\'ari}
\affiliation{Konkoly Observatory, MTA CSFK, H-1121 Budapest, Konkoly Thege M. \'ut 15-17, Hungary}

\author[0000-0002-2943-5978]{Lidia van Driel-Gesztelyi}
\affiliation{Konkoly Observatory, MTA CSFK, H-1121 Budapest, Konkoly Thege M. \'ut 15-17, Hungary}
\affiliation{University College London, Mullard Space Science Laboratory, Holmbury St. Mary, Dorking, Surrey, RH5 6NT, UK} 
\affiliation{LESIA, Observatoire de Paris, Universit\'e PSL, CNRS, Sorbonne Universit\'e, Universit\'e de Paris, 5 place Jules Janssen, 92195 Meudon, France}

\author{Attila Mo\'or}
\affiliation{Konkoly Observatory, MTA CSFK, H-1121 Budapest, Konkoly Thege M. \'ut 15-17, Hungary}

\author[0000-0001-5449-2467]{Andr\'as P\'al}
\affiliation{Konkoly Observatory, MTA CSFK, H-1121 Budapest, Konkoly Thege M. \'ut 15-17, Hungary}



\begin{abstract}
We analyze the light curve of the M5.5 dwarf Proxima Centauri obtained by the TESS in Sectors 11 and 12. In the $\approx 50$ day-long light curve we identified and analyzed \FLARENUMBER{} flare events. The flare rate was 
\FPeval{\result}{round(\FLARENUMBER/ (\OBSTIME /24.),2)}\result{}
events per day, in total, \FLARERATIO{}\% of the data was classified as flaring. 
The estimated flare energies were in the order of $10^{30}-10^{32}$ ergs in the TESS passband  ($\approx4.8\times$ higher in bolometric energies, but in the same order of magnitude). Most of the eruptions appeared in groups. Two events showed quasi-periodic oscillations during their decay phase with a time scale of a few hours, that could be caused by quasiperiodic motions of the emitting plasma or oscillatory reconnection.
From the cumulative flare frequency distribution we estimate that superflares with energy output of $10^{33}$ ergs are expected to occur three times per year, while a magnitude larger events (with $10^{34}$ ergs) can occur every second year. This reduces the chances of habitability of Proxima Cen b, although earlier numerical models did not rule out the existence of liquid water on the planetary surface.
We did not find any obvious signs of planetary transit in the light curve.

\end{abstract}



\section{Introduction}

At present, low-mass, cool M dwarfs are the prime targets of planet searches, since
 the habitable zone is much closer to the central object in cool stars than in the case of a solar-like star; thus, detecting a possibly habitable Earth-like planet is easier.
Proxima Centauri, the nearest star to our Sun, was an evident subject of such a research \citep[e.g.][]{1999AJ....118.1086B,2008A&A...488.1149E} -- recently, \cite{2016Natur.536..437A} reported the discovery of Proxima b, a $1.27M_\oplus$ planet orbiting within the habitable zone. The late spectral type (M5.5V) of the central star and the magnetic activity associated with it, however, could  pose a threat to habitability.

The role of magnetic activity on planetary habitability is a currently actively studied field of research (see \citealt{2007AsBio...7...85S,2007AsBio...7...30T,2010ARA&A..48..631S} and references therein).
Flares, coronal mass ejections, and associated high-energy radiation and particles can have a serious impact on the environment by gradually evaporating planetary atmospheres
\citep{2007AsBio...7..167K,2008SSRv..139..437Y,2017A&A...608A..75C}. 
A recent study of \cite{2019A&A...623A..49V}, however, suggests that  radiation effects would play the main role compared to coronal mass ejections in exoplanetary atmosphere evolution.
Such frequent, high-energy events could cause the planetary atmospheres to be continuously altered, which is disadvantageous for hosting life (see  \citealt{2017ApJ...841..124V,rachael-trappist1} and references therein), however, a strong enough planetary magnetic field could provide some protection from these harmful effects \citep{2013A&A...557A..67V}.

Proxima Centauri is one of the few currently known ultracool objects that host an Earth-mass planet in its habitable zone 
beside the TRAPPIST-1 system  \citep{trappist1},
Teegarden's star \citep{teegarden-carmenes}, 
and the recently discovered GJ 1061 \citep{2019arXiv190804717D} 
and TOI-270 \citep{2019NatAs.tmp..409G} systems -- 
and is therefore an important proxy for understanding planet formation and evolution around ultracool dwarfs. 
Proxima Cen is the third, distant member of the $\alpha$~Cen system consisting of two components of nearly solar mass A (G2V) and B (K1V) which form a visual binary provided all three stars were formed together which is not yet fully settled (cf. \citealt{2018MNRAS.473.3185F}). Stars A and B have rotational periods and magnetic activity cycle lengths resembling to the solar values: 17.5~days and $>20$~yrs for component A, and 36.23~days and 8.9~yrs for component B, respectively. \cite{2018AA...615A.172M} found an age of $\approx$6~Gyr for $\alpha$~Cen A and B from abundance indicators, which compares well to the possible age of Proxima Cen (about 5.1~Gyr) following from the rotation--age relation of \cite{2018RNAAS...2a..34E} with the rotational period of 83 days. The stellar system containing Proxima Cen is possibly older than the Sun.
Proxima Centauri has a moderately strong magnetic field of $\approx450-750$\,G \citep{magneticfield}.
Its rotation period was estimated to be $P=83.5$ days from HST data \cite{rotation1} and $P=82.5$ days using ASAS photometry \citep{rotation2}. Despite being a slowly rotating, fully convective ultracool dwarf, it shows an activity cycle of 
$\approx7$ years (see \citealt{cycles2} and references therein), although this would be expected for faster rotating stars of solar-like structure (see \citealt{2013AN....334..972V,cycles2} and references therein).
Parameters of Proxima Centauri and its planet are summarized in Table \ref{tab:params}.

\begin{table}
\centering
\caption{Parameters of Proxima Centauri system}
\label{tab:params}
\begin{tabular}{lll}
\hline\hline
\multicolumn{2}{c}{Parameters of Proxima Centauri}&Reference\\
\hline
Mass            & $0.120\pm0.003M_\odot$    &\cite{ribas}               \\
Radius          & $0.146\pm0.007R_\odot$    &\cite{ribas}               \\
Luminosity      & $0.00151\pm0.00008L_\odot$&\cite{ribas}               \\
$T_\mathrm{eff}$& $2980\pm80$\,K               &\cite{ribas}               \\
$\log g$        & $5.02\pm0.18$             &\cite{passegger}           \\
$[$Fe/H$]$      & $−0.07 \pm 0.14$          &\cite{passegger}           \\
Parallax        &$768.5\pm0.2$\,mas           &\cite{2018yCat.1345....0G} \\
Age             &$\approx 6$\,Gyr             &\cite{2018AA...615A.172M} \\
Magnetic field strength& 450--750\,G ($3\sigma$)   &\cite{magneticfield}       \\
Rotation period & $\approx83$\,days          &\cite{rotation1,rotation2} \\
Activity cycle length&$\approx7$\,years     &\cite{cycles2}               \\
Habitable zone range &$\approx 0.0423-0.0816$\,AU&\cite{2016Natur.536..437A}  \\
Habitable zone periods&$\approx 9.1-24.5$\,days&\cite{2016Natur.536..437A}  \\
\hline
\multicolumn{2}{c}{Parameters of Proxima Centauri b}&\\
\hline
Orbital period  &   $11.186\pm0.001$\,days   &\cite{2016Natur.536..437A}  \\
Semi-major axis, $a$& $0.0485\pm0.05$\,AU     &\cite{2016Natur.536..437A}  \\  
Minimum mass    & $1.27\pm0.19 M_\oplus$    &\cite{2016Natur.536..437A}  \\  
Equilibrium temperature & $230\pm10$\,K       &\cite{2016Natur.536..437A}  \\  
\hline
\end{tabular}
\end{table}

The flaring activity of Proxima Centauri was detected in multiple wavelengths from the mm wavelength to the X-ray regime \citep{mmflare,xrayflares,multiwavelength}.
Based on high-resolution optical spectra \cite{flare-photosphere} concluded that the atmospheric structure of the star is rather complicated, consisting  of a normal M5 dwarf photosphere, an extended hot envelope and hot stellar wind with a typical velocity of $V_r=30$\kms{} that yields a minimum mass loss of $\dot M = 1.8\times 10^{-14}M_\odot$ per year. 
The chromosphere layers were suggested to be heated by flares \citep{flare-temporal}.
Flare events were reported also using photometric observations: 
\cite{davenport} analyzed MOST observations of 37.6 days, and found 66 flare events in white light with energies of $10^{29}-10^{31.5}$ ergs.
In 2016 the Evryscope \citep{evryscope} detected an eruption in $g'$ band that caused $\approx 68\times$ flux increase -- a naked-eye superflare --, that was estimated to have a bolometric flare energy of $10^{33.5}$ ergs, and several other flares in the range of $10^{30.6}-10^{32.4}$ ergs \citep{superflare-evryscope}; while 
in 2017 another superflare was detected in $i'$ passband with at least 10\% flux increase and an energy output of $>10^{33}$ ergs \citep{superflare}.
In this paper, we analyze photometric observations of Proxima Centauri obtained by the TESS (Transiting Exoplanet Survey Satellite).

\section{TESS observations and data reduction}
\begin{figure}
\plotone{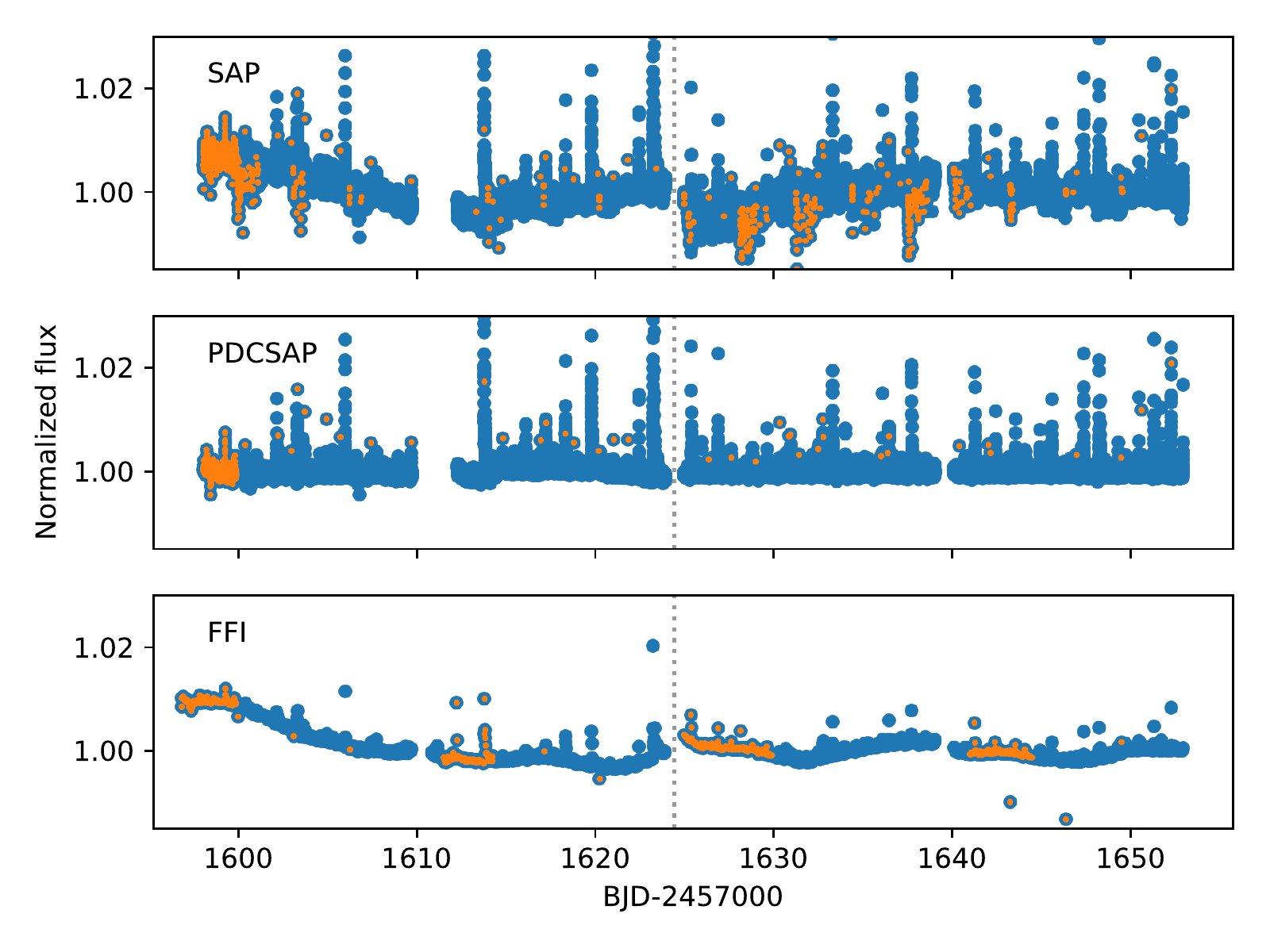}
\caption{Comparison of the different photometric methods, from top to bottom: simple aperture photometry (SAP), Pre-search Data Conditioning (PDCSAP), long-cadence data from full-frame images (FFI). Data points flagged as having quality problems are overplotted in orange. Dotted vertical line separates data from Sectors 11 and 12. The plots do not show the full range of the light curves.}
\label{fig:comp_lc}
\end{figure}

Proxima Centauri was observed by TESS in Sectors 11 and 12. Calibrated, short-cadence (2-min) data were downloaded from the MAST site%
\footnote{\url{https://mast.stsci.edu}}. 
We decided to use Simple Aperture Photometry (SAP) data, since in the Pre-search Data Conditioning (PDC) light
curve instrumental and astrophysical signatures are removed to better isolate transits and eclipses, which is the primary scientific goal of the mission. The PDCSAP data show a somewhat lower scatter level, and it also removes a long-term trend from the light curve, that could be compatible with the $\approx 80$ day-long rotation period. Since the length of the dataset (53 days) is shorter than the rotation period, the validity of this trend cannot be safely confirmed, however, this does not influence our further analysis. Oscillations on hourly time scale  are present in both data sets (see Sect. \ref{sect:oscillations} for details and Fig. \ref{fig:bumps} for a comparison of the two light curves).

From the light curve we excluded points where the quality flags indicated problems
using bit-wise AND operation with the binary mask 101010111111 
as suggested by the TESS Data Product Overview%
\footnote{\url{https://outerspace.stsci.edu/display/TESS/2.0+-+Data+Product+Overview}}.

As a validity check, we also performed a separate photometry on the 30-min cadence full-frame images (FFI) to see if there were some serious issues with the automated pipeline. 
This photometry is based on differential imaging algorithms, implemented by various tasks of the FITSH package \citep{pal2012}. In this processing scheme, smaller, $96\times96$ pixel sized image stamps are trimmed from the original full frame image (FFI) series, centered at the source. Columns with various artifacts  were masked in order to avoid in the derivation of the best-fit convolution transformation. 
The most prominent ill-calibrated block of CCD columns were associated with the blooming of the image of $\alpha$ Cen where charges have bloomed into the smear region of the CCD -- therefore, the automatic calibration procedure underestimated the pixel values at all of the affected columns.  This convolution transformation accounts for all linear instrumental effects, including smear, spacecraft jitter, gradual offset due to differential velocity aberration and background variations (due to stray light). Once these convolution kernel parameters are determined, differential and absolute fluxes were obtained by Equations (82) and (83) of \cite{pal2009}, respectively. Similarly to the short-cadence data, full frame images are also flagged in a similar bit-mask scheme and photometric points corresponding to frames with unexpected flags are removed from further analysis.
The resulting normalized light curves from the SAP, PDCSAP and FFI data are compared in Figure \ref{fig:comp_lc}. The long-term trend is visible in both the SAP and FFI light curves, suggesting that this is a physical phenomenon rather than an instrumental effect. 

\cite{2016Natur.536..437A} predicted a transit depth of $\approx 0.5\%$ (5\,mmag) with a geometric transit probability of $\approx 1.5\%$ and $P=11.186$ days. A search of \cite{notransit} ruled out transits with this period, and events shorter than 5 days with depths over 3\,mmag. Although this is outside the main scope of this current paper, we did a quick period analysis to search for a possible transit, as an event of a few mmag depth should be easily noticeable (the precision of the FFI light curve is about 0.1\,mmag), however, the predicted depth is not present in the data series.

\section{Flare detection and flare energies}
\label{sect:energies}
\begin{figure}
\plottwo{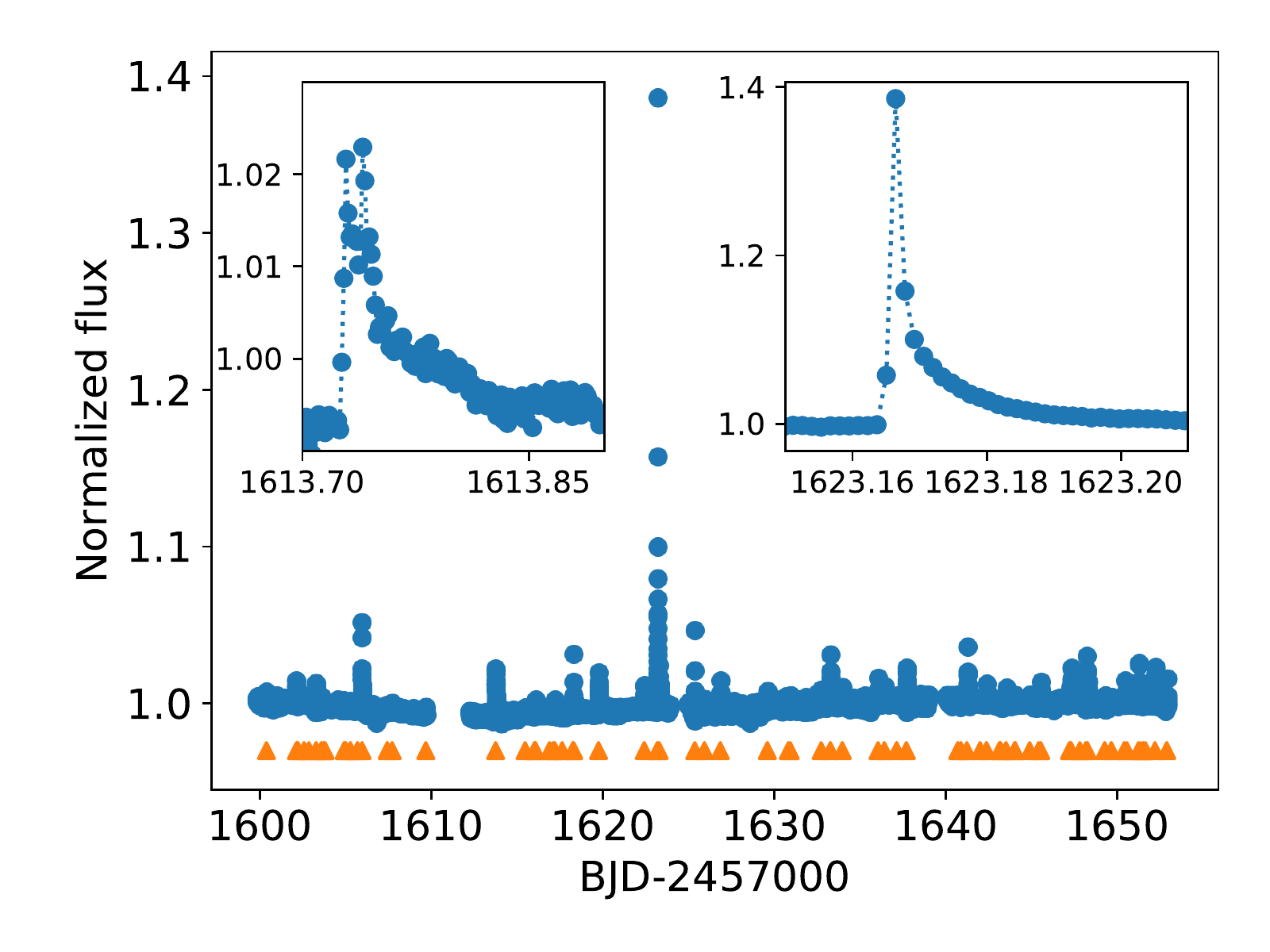}{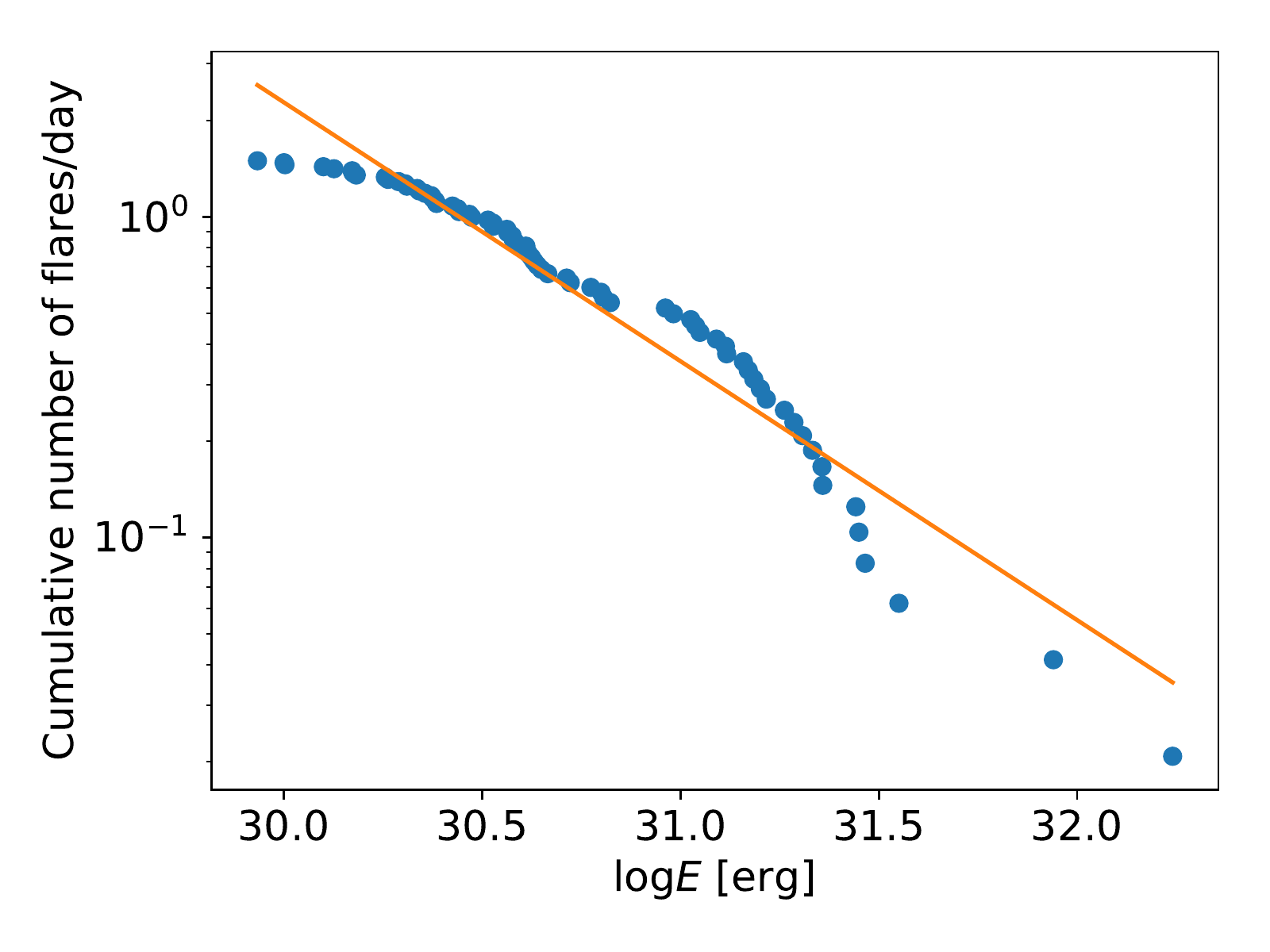}
\caption{Left: TESS light curve (normalized SAP flux) of Proxima Centauri. Two insets show zoomed-in light curves of the largest flare events. Triangles mark the times of the identified flares. Right: Cumulative flare frequency distribution fitted by a linear function.}
\label{fig:flatwrm}
\end{figure}

Flare identification was done by visual inspection, in order to safely select also smaller events, that automated algorithms tend to miss.
To test the detection threshold of the visual inspection, we prepared a test dataset with a scatter typical to the TESS data. To this artificial light curve 25 flares were added with random lognormal amplitude distribution (similar to the actual observations), with amplitudes ranging between $\approx 0.5-3\%$ flux increase. This light curve was then analyzed in the same manner as the original TESS light curve. We found that the detection limit of our method is at roughly 0.001--0.002 increase in normalized intensity.
Of the \FLARENUMBER{} detected events, a sample is shown in Figs. \ref{fig:flatwrm} and \ref{fig:flarezoo}. 
The light curves of the events were fit using the flare template of \cite{davenport-model}. The fits to the normalized light curves yielded amplitudes in the range of $0.0007-0.2$ with a mean and median of 0.027 and 0.015; and full-width at half maximum (FWHM) values -- describing roughly the time scale of the events -- in the range of $0.0012-1.74$ days, with mean and median values of 0.05 and 0.005, respectively. Note, that the analytic models do not always fit well to the observations (especially in the case of small or complex events, or poorly sampled data), but these numbers can help to give an idea about the amplitude and time scale of the flares.

The flare energy estimation was done by integrating the normalized flare intensity during the event (see e.g. \citealt{flatwrm}):
\begin{equation}
\varepsilon_f = \int\limits_{t_1}^{t_2} \left( \frac{I_{0+f}(t)}{I_0}-1\right),
\end{equation}
where $t_1$ and $t_2$ are the beginning and end times of the event, and $I_{0+f}$ and $I_0$ are the intensities with and without a flare. This value gives the relative flare energy, or equivalent duration. To get the energy in the observed bandpass ($E_f$), we have to multiply this by the quiescent stellar  luminosity ($L_\star$):
\begin{equation}
E_f = \varepsilon_f L_\star.
\end{equation}
We estimated the quiescent luminosity using the spectrum of Proxima Centauri compiled by \cite{ribas}. This spectrum was convolved with the TESS response function 
and integrated over wavelength to obtain the observed quiescent luminosity 
yielding $L_\star=9.915\times10^{29}$ \ergs{} for the TESS bandpass%
\footnote{available e.g. at the TESS Science Support Center \url{https://heasarc.gsfc.nasa.gov/docs/tess/the-tess-space-telescope.html}}.
The resulting flare energies are plotted in Figure \ref{fig:flatwrm} and summarized in Table \ref{tab:energies} and \ref{tab:energies2}.

If $dN$ is the number of flares in the energy range $E+dE$, $dN(E)$ can be written as
\begin{equation}
 dN(E)\propto E^{-\alpha}dE
\end{equation}
(see e.g. \citealt{hawley} and references therein).
The cumulative flare frequency distribution can be expressed in logarithmic form by integrating as
\begin{equation}
\log \nu = a + \beta \log E,
\end{equation}
where $\nu$ is the cumulative frequency of the flares with the given energy larger than $E$, with $\beta = 1-\alpha$ \citep{gizis}. To describe the characteristics of a flaring star, this distribution is usually fitted by a linear function, that yields the slope $1-\alpha$. The best fit gives $\alpha=\ALPHALINEAR$ (see Fig. \ref{fig:flatwrm}). The given error is the formal error of the linear fit, the true uncertainty is probably a magnitude larger. 
Following \cite{gizis}, an alternative approach is using an unbiased maximum likelihood estimator (corrected for the small sample size):
\begin{equation}
\alpha -1 = (n-2) \left[ \sum_{i=1}^n \ln \frac{E_i}{E_\mathrm{min}}\right]^{-1},
\end{equation}
that yields $\alpha=\ALPHAME$ .
This method has the advantage of being independent of the energy range chosen for the fit, that can significantly change the result.
From MOST photometry \cite{davenport} concluded a similar result of $\alpha=1.68$.

\begin{figure}
\plotone{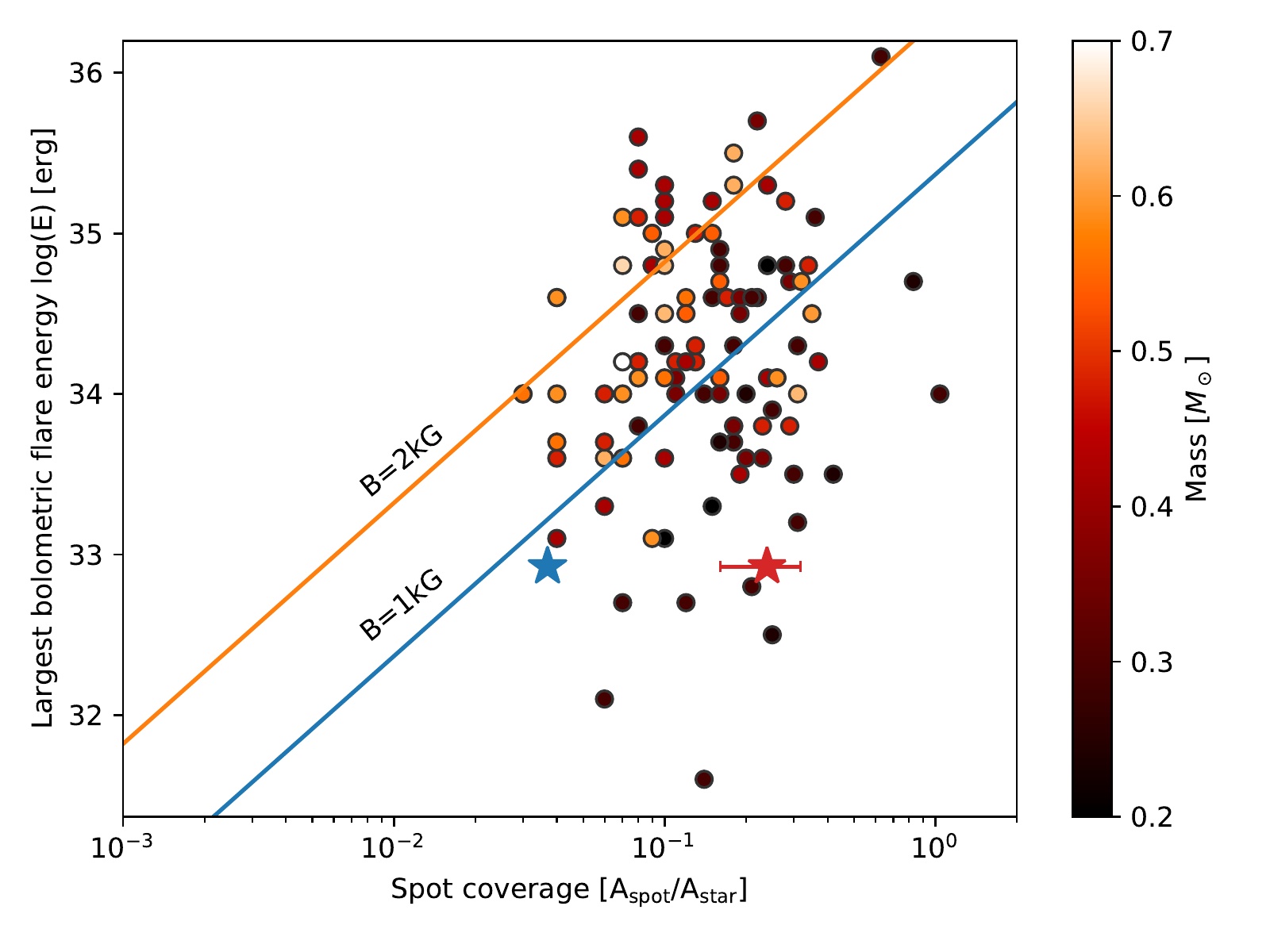}
\caption{Flare energies as a function of spot area as fraction of the projected visible surface. Red dots show data from the Evryscope project \citep{2019arXiv190710735H}, blue  and red stars show the largest observed flare on Proxima Cen in the TESS data using the method of \cite{2019arXiv190710735H} and an estimation based on the measured global magnetic field, respectively. In the latter case (red star) the error bar corresponds to the $3\sigma$ uncertainty of the magnetic field \citep{magneticfield}. Solid lines show minimum spot coverage needed to generate flares at observed energies with 1 kG and 2kG field strength. The energy released by the event on Proxima Cen falls between that of solar and stellar flares (cf. \citealt{2019ApJ...876...58N}).}
\label{fig:flarearea}
\end{figure}
According to \cite{2013PASJ...65...49S, 2019ApJ...876...58N} the bolometric energy released by a flare ($E_\mathrm{flare}$) and 
the area of the smallest spot that could
produce such an event ($A_\mathrm{spot}$) are related as
\begin{equation}
E_\mathrm{flare}\approx fE_\mathrm{mag}\approx \frac{B^2}{8 \pi}A^{3/2}_\mathrm{spot},
\end{equation}
where $B$ is the magnetic field strength, $f$ is the fraction of magnetic energy $E_\mathrm{mag}$ that can be released as a flare. 
To estimate the minimum areas for the flares observed by TESS, we converted the derived energies to bolometric ones. Following \citet{gunther2019}, in 
this calculation adopting 9,000\,K blackbody radiation model for flare emission we obtained
 $\sim$4.8$\times$ larger bolometric energy release than that in the
TESS-band. By applying the above relation then we found that the observed flares come from an area $A_{spot}$ at least \FLAREAREA{} of the projected visible stellar surface, using the bolometric energy of the largest observed flare a mean magnetic field strength $B$ of 450--750\,G, and supposing that the fraction of magnetic energy released as a flare ($f$) is 10\% (cf. \citealt{2013PASJ...65...49S}). This calculation overestimates the spot area, as the magnetic field strength in the active nest is higher than the measured global field. Following the method of \cite{2019arXiv190710735H} and \cite{2019ApJ...876...58N} we can give another estimation of the spot area. First, we estimate the temperature of the starspots from the effective temperature:
\begin{equation}
    \Delta T(T_\mathrm{star}) = T_\mathrm{star} - T_\mathrm{spot} = 
    3.58 \times 10^{-5} T_\mathrm{star}^2 + 0.249T_\mathrm{star} - 808,
\end{equation}
where $T_\mathrm{star}$ and $T_\mathrm{spot}$ are the temperatures of the unspotted and spotted photosphere. With normalized amplitude of the light variation caused by the spottedness ($\Delta F/F$) the spot area can be estimated based on the TESS light curve:
\begin{equation}
    A_\mathrm{spot} = \frac{\Delta F }{F} A_\mathrm{star} 
    \left[ 1- \left( \frac{T_\mathrm{spot}}{T_\mathrm{star}} \right)^4 \right]^{-1}
\end{equation}
yielding a spot area of $\approx4\%$.
This result agrees well with 
results of \cite{2019arXiv190710735H} (see Fig. \ref{fig:flarearea}) plotting the most powerful flare from Table~\ref{tab:energies}. 
The event on Proxima Cen falls between the rarely populated area of solar and stellar flares (cf. \citealt{2019ApJ...876...58N}). We stress, that the mass of Proxima Cen is lower than any of the stars in that sample (see Table~\ref{tab:params}), that falls between (0.2--0.7$M_\odot$).

TESS was observing Proxima Centauri for a total of \OBSTIME{} hours. During this time we detected \FLARENUMBER{} flare events, thus the total flare rate was 
\FPeval{\result}{round(\FLARENUMBER/ (\OBSTIME /24.),2)}\result{}
flares per day (%
\FPeval{\result}{round(\FLARENUMBER/ (\OBSTIME),3)}\result{}
flares per hour). In total, \FLARERATIO \% of the total observing time was classified as flaring -- Sector 11 data being somewhat less active than that of Sector 12. 
These values are similar to those found by \cite{davenport} using MOST observation: they measured 8.1 flares per day in 2014 and 5.7 flares per day in 2015 (a total of 7.5\% of the total observing time). 
We note, that these numbers could be biased compared to a fast-rotating star -- in that case, it is possible that a flare occurring on the opposite hemisphere can be detected as the star rotates within the time scale of the eruption, while in the case of Proxima Cen all the TESS observations do not cover even a whole rotation.
According to \cite{cycles2} the magnetic cycle length is $\approx 7$ years, with an activity minimum in 2012/2013. This means that the observation of the MOST satellite was done halfway to the activity maximum, while the TESS light curve was obtained close to the minimum. Interestingly there is no significant change in the relative flaring activity across these epochs. After finishing its primary mission, the TESS could revisit Proxima Cen, and using those observations we could learn more about the long term changes of the flaring activity of the object.

The flaring rate of Proxima Cen is similar to most of those observed on well-known flare stars, e.g. AB~Dor: 0.5 \citep{1990A&AS...85.1127J} or EV~Lac: 4.2 \citep{1987A&A...177..201D} flares per day, which have rotational periods from about 0.5 to a few days. 
Recently, \cite{2019ApJS..241...29Y} rigorously analyzed the K1 {\it Kepler} data for flaring stars. According to their Table~1., the highest flare frequencies -- only for a few stars -- are around 0.35 flare per day, and these stars have rotational periods of a few days. 
Looking at the slowest rotators, the dwarf flare star KIC~11027877 in \cite{2019ApJS..241...29Y} with the longest rotational period  of 61.22 days has a flare frequency of 0.02 flares/day, and another dwarf (KIC~9203794) with a rotation of 52.27 days has 0.1 flares/day.
Comparing Proxima~Cen to these results, it seems that it has a very high flaring rate with an even slower rotational period, though the connection between the rotational rate (decreasing with age) and flaring activity would suggest otherwise \citep{2019ApJ...871..241D}.

\section{Observed bumps during flare decay}
\label{sect:oscillations}

\begin{figure}
\plottwo{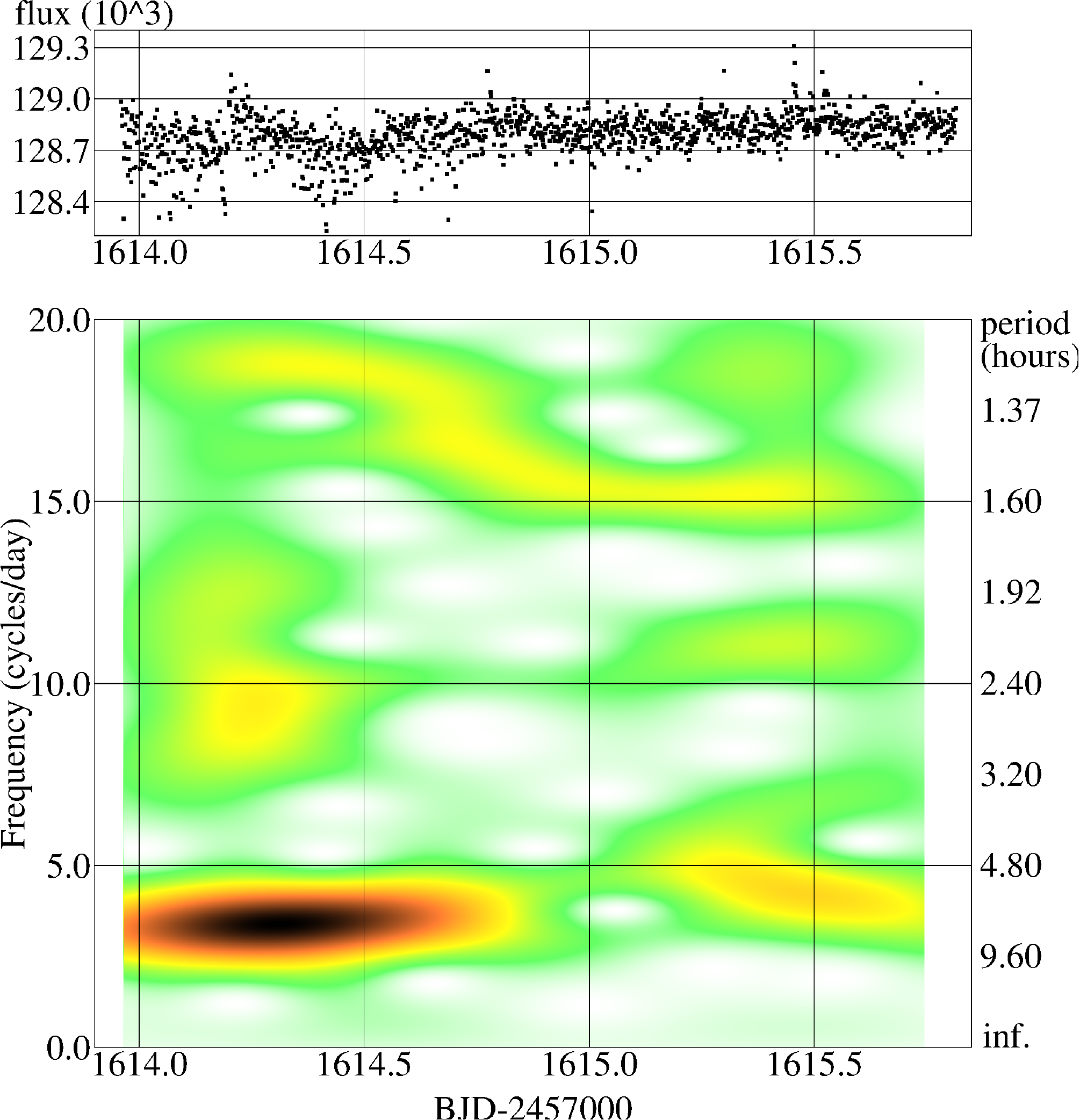}{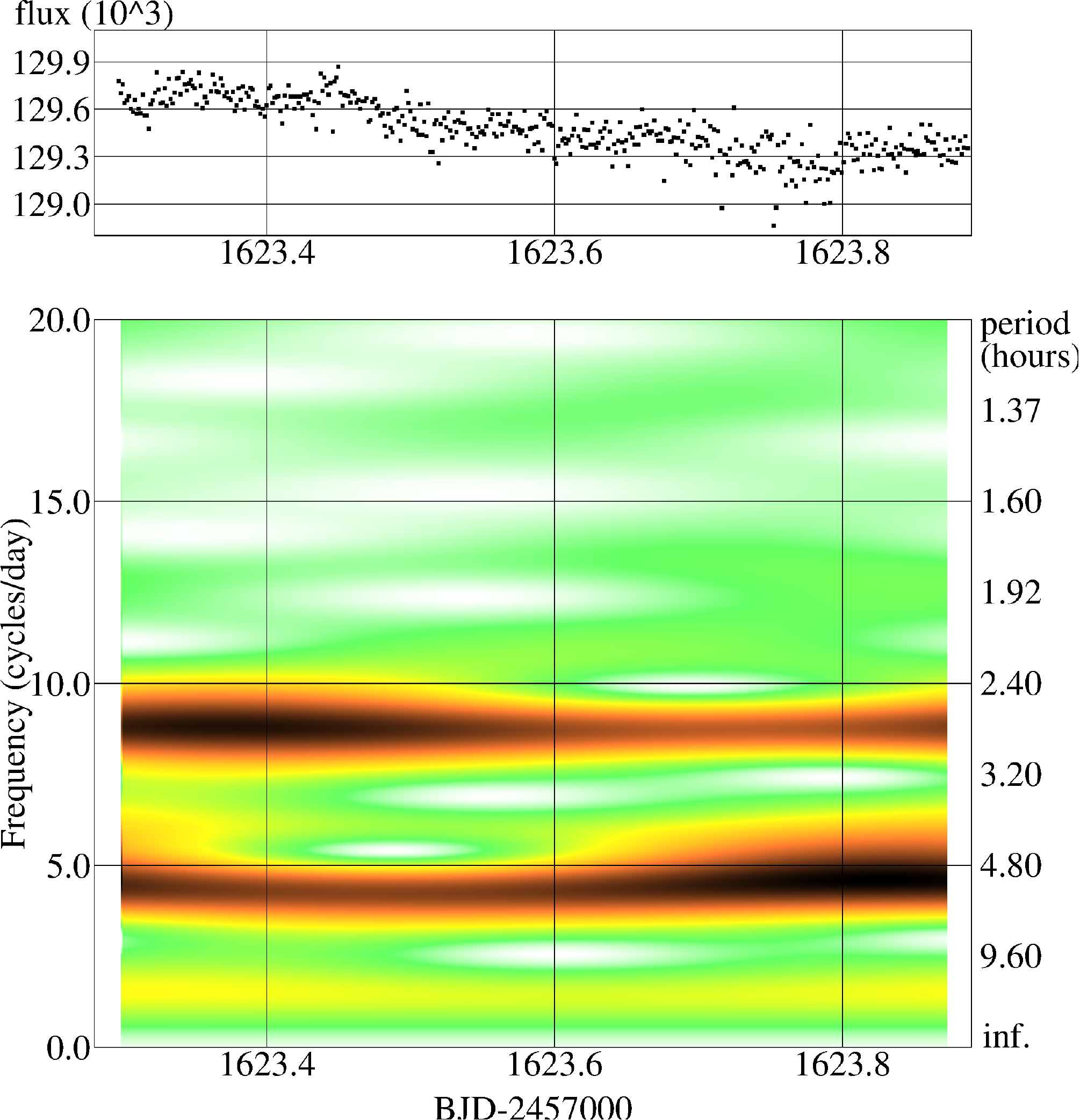}
\caption{SAP fluxes after the two flares having post-flare bumps (top) and their short-term Fourier-transform (STFT, bottom). {\it Left:} time scale of the bumps is about 6.5 hours, {\it right:} repeating bumps of 2.7 and 5.4 hours. 
The black color marks the highest amplitude of the signal in the plot, while orange, yellow and green show 50\%, 20\% and 5\% of that with continuous transition.
See text for more.}
\label{fig:stft}
\end{figure}

The flares of Proxima Cen appear mostly in groups of two to many flares. From the \FLARENUMBER{} eruptions we find only a few definitely single events. The background of the multiplicity can be twofold: either the flares originate from the same activity source or, especially in case of very high flare frequency, are just coincidences (see  \citealt{hawley} for details).

From the observed flares we find two events which have a long decay phase. The first one lasts for over 1.5 days, and the second one for at least several hours (see the upper panels in Fig. \ref{fig:stft} and Fig. \ref{fig:bumps}) and is a double event with a smaller secondary eruption. In the latter case the full decay is not seen, as the observations in Sector 11 finished before the star could return to its quiescence. The observations of these two flares were cut after the faster decay phase and the remaining data from the slower decay were analyzed using the time--frequency method of short-term Fourier transform \citep{2009A&A...501..695K}.
The lower panels in Fig. \ref{fig:stft} show the results of this time--frequency analysis. 
A pre-emphasis filter has been applied to suppress all the periodicities longer than 8.2 hours, since the amplitude of the decay itself is higher in the second case than the amplitudes of the bumps (for the details see  \citealt{2009A&A...501..695K}, Sect. 2.). For the first event (left plot in Fig. \ref{fig:stft}) we find repeated bumps on the time scale of about 6.5 hours lasting for nearly a day with decreasing amplitude, and then damping until the quiescence is reached (total analyzed data is about 1.85 days = 44.5 hours long). The second event is more energetic (see also Table \ref{tab:energies}), and its decrease  shows clear repeating bumps on the time scales of 2.7 and its double, 5.4 hours; the shorter period has higher amplitude in the beginning of the decrease, while the longer one is stronger at the end of the observations, when very probably the decay is still not finished (total analyzed data is about 0.6 days = 14 hours long).

Flare oscillations were observed previously on Proxima Cen in X-rays \citep {2016ApJ...830..110C}  and on several other stars in NUV \citep{2018MNRAS.475.2842D}. This latter paper also uses time--frequency method (wavelet) to find the oscillation time scales. All these derived oscillation time scales for flare stars (including Proxima Cen) are in the order of a few to a few tenths of seconds (between 20--120 sec), i.e., much shorter than those we find for the bumps in the TESS wavelength range (ro2ughly white light). It is likely that the origins of these features are different.

By analyzing Kepler light curves of different types of flaring stars from dwarfs to giants, \cite{2015MNRAS.450..956B} and also \cite{2016MNRAS.459.3659P} found 
quasi-periodic bumps with the characteristic periods of an hour (between 3--110 minutes).
These values are the in same order as our result for the Proxima bumps. The characteristic time scales of the bumps can be very different on one star at different times, and the periods of the bumps do not show correlation of the physical parameters of the stars (cf. \citealt{2016MNRAS.459.3659P}, their Fig. 2.). We note, that Proxima Cen is much different from the stars observed by {\it Kepler} and studied by \cite{2015MNRAS.450..956B} and \cite{2016MNRAS.459.3659P}. Its rotational period ($\approx$83 days) is about twice as long as that of KIC~2852961 (35.5 days, \citealt{2016MNRAS.459.3659P}) and KIC~5952403 (45.28 days, \citealt{2015MNRAS.450..956B}), which have bumps during their flare decrease, both of them are giants, while Proxima Cen is a low-mass dwarf star (Table~\ref{tab:params}).

The oscillatory pattern we find in the decay phase of these two long-duration flares on Proxima Centauri appears to be very similar to quasi-repetitive  patterns with periods ranging from sub-second to several minutes observed in solar and stellar flares which are called quasi-periodic pulsations (QPP; e.g. \citealt{2016SSRv..200...75N}). QPPs have been observed over the electromagnetic spectrum from radio through optical to X-rays indicating that QPPs affect all layers of the solar or stellar atmosphere from the photosphere to the corona.
Statistical studies seem to indicate that QPPs are not a rare phenomenon. Although not all flares show observable QPPs, larger flares tend to have such oscillatory pattern. For example, \cite{2015SoPh..290.3625S} found that about 80 \% of X-class flares of the present solar cycle 24 had detectable QPPs.

Although there is still no consensus about the exact underlying physical mechanisms, there are several plausible candidates, which can be broadly divided in two categories \citep{2018SSRv..214...45M}: (quasi-)periodic motions of the emitting plasma around an equilibrium, connected with the competition between inertia and an effective restoring force (incl. MHD oscillations) and self-oscillatory mechanisms (e.g. load-unload models and relaxation processes in which a steady inflow of magnetic flux towards a reconnection site could result in repetitive magnetic reconnection (so-called “magnetic dripping”, \citealt{2010PPCF...52l4009N}).  Another credible model in this category is oscillatory reconnection \citep{2009A&A...494..329M,2009A&A...493..227M,2012A&A...544A..24T}, which -- like other models that invoke magnetic reconnection -- gives a natural explanation of the QPPs' multi-wavelength nature. In the oscillatory reconnection model there is a competition between thermal-pressure and magnetic-pressure gradients, each successively overshoot the equilibrium created by the other. The resulting successive bursts of reconnection decrease in power as both pressure gradients are decreasing with time, allowing the system to approach an equilibrium state. In the model by \cite{2009A&A...494..329M} oscillatory reconnection occurred between a newly emerging magnetic flux and its locally open magnetic field environment. Emerging flux is being one of the main flare triggers, the proposed mechanism appears plausible in the QPP-like oscillations we observe in the decay phase of two powerful flares on Proxima Centauri.

\section{Implications on habitability}
\label{sect:habitability}
Proxima Cen b is a $1.27M_\oplus$ planet, that orbits roughly at 1/20th Sun--Earth distance to its host with an orbital period of roughly 11 days. The age of the system ($\approx 6$Gyr) would also make it a good target for the search of life. 
Numerical models suggest that the planet could have lost about an ocean's worth of water due to the early irradiation in the first 100--200 million years of its life
\citep{2016A&A...596A.111R,2016A&A...596A.112T}, although the amount of initial water on the planet is unknown. After this period Proxima Cen b could either end up as a dry, atmosphereless planet by further loss its atmospheric gases, or
it could keep most of the atmosphere preserving liquid water on the surface. In the latter scenario the authors concluded that liquid water may be present over the surface of the planet in the hemisphere of the planet facing the star, or in a tropical belt.
According to these models, it cannot be ruled out that Proxima Cen~b could be considered a viable candidate for a habitable planet -- this makes the effect of external factors, like flaring activity of the host star, even more interesting.

Unusually energetic flare events are often referred as 'superflares'. The exact threshold for naming an eruption superflare is somewhat arbitrary in the literature, but \cite{superflare} suggested  $10^{33}$ ergs -- roughly ten times the energy output of the Carrington event on the Sun -- as a reasonable threshold. Currently two of such events were observed on Proxima Cen: \cite{superflare-evryscope} observed an eruption with a flux increase of $\approx 68$ and a bolometric energy of $10^{33.5}$ ergs; and \cite{superflare} found an other event with an estimated energy  in the order of $\approx 10^{32}-10^{33}$ ergs in Sloan $i'$ band.
According to \cite{davenport}, the frequency of such outbreaks is $\approx 8$ times per year. From the TESS data we estimate roughly 3 superflares per year with an energy of $10^{33}$ ergs, and about one event in every two year with an energy output of $10^{34}$ ergs (note, that the TESS data were obtained roughly at the minimum of the magnetic activity cycle of Proxima Cen).  
These numbers are much higher, than the estimated superflare frequency for G-type stars \citep{2013ApJS..209....5S}, roughly one per thousand year ($10^{34}-10^{35}$ ergs), therefore, such events could have a more serious effect on their surroundings as in the case of solar-like stars, especially since the planets orbit much closer to their hosts in late-type stars as in solar-like objects.

The effects of flares on exoplanetary habitability is strongly debated (cf. \citealt{lingam}).
UV radiation can modify, ionize, and even erode planetary atmospheres over time -– leading to the photodissociation of important molecules such as water and ozone \citep{2007AsBio...7..167K,2008SSRv..139..437Y}.
On the other hand, planets orbiting M dwarfs may not receive enough UV flux for abiogenesis (i.e, the process by which life can arise from non-living simple organic compounds), which could be remedied by frequent flares \citep{2017ApJ...843..110R}.
While \cite{2010AsBio..10..751S} showed that single, large flare events do not threat habitability around M-dwarfs, strong, frequent flares, can cause the planetary atmospheres to be continuously altered, making them less suitable for habitability 
(see  \citealt{2017ApJ...841..124V,rachael-trappist1}, and references therein).
%

Currently there are only a few M-dwarf systems that are known to host Earth-like planets. These planetary systems are much tighter than our Solar System -- e.g., the planets around TRAPPIST-1 orbit between 0.01--0.06\,AU around their host, and the semi-major axis of Proxima Centauri b is 0.05\,AU (roughly at the distance of TRAPPIST-1 g). At this distances, both the radiation and the particle flux for such planets are much higher than for Earth -- roughly 400 times higher for Proxima Centauri b compared to what the Earth receives when the Sun emits a flare of the same energy. 
Therefore, the activity of the central star can be more harmful to its habitable-zone planets in the case of M-dwarfs. In two of the five such currently known systems -- Proxima Centauri and TRAPPIST-1 -- the magnetic activity of the host seems to pose a serious threat to the habitability of their planets. 
Teegarden's star \citep{teegarden-carmenes}, with an estimated age of $>8$ Gyrs, has two known $1.3 M_\oplus$ planets, 
and \cite{2019ApJ...880L..21W} found that surface liquid water could be present on both planets for a wide range of atmospheric properties, making these attractive targets for biosignature searches.
The host star, however, is also known to flare \citep{teegarden-carmenes}, but currently not much is known about its activity in detail. 
There are, however, two recently discovered systems, that might provide a friendlier environment for life.
GJ 1061 hosts a $1.5 M_\oplus$ planet with the central star showing only occasional small flares \citep{2019arXiv190804717D}, 
while TOI-270 is a nearby, quiet M-dwarf with a super-Earth ($1.25R_\oplus$) and two sub-Neptunes ($2.42$ and $2.13R_\oplus$). TOI-270 did not show any signs of rotational variability or flares during the photometric observations, and seems to have a low activity according to H$\alpha$ measurements  as well \citep{2019NatAs.tmp..409G}.
These possibly older systems (GJ 1061 is $>7$Gyrs old, TOI-270 has currently no age estimation) could be good candidates for detailed habitability studies, if the atmospheres of their planets could survive the early active phase of the host star; or if the planets could acquire a secondary atmosphere at later stages of their lives, e.g. due to late bombardment \citep{2018MNRAS.479.2649K,2019MNRAS.487.2191D} or outgassing \citep{2019A&A...625A..12G}.

\section{Summary}
\begin{itemize}
    \item In the 53 day-long light curve of Proxima Cen obtained by TESS in Sectors     11 and 12 we found \FLARENUMBER{} flare events;
    \item The flare rate was
        \FPeval{\result}{round(\FLARENUMBER/ (\OBSTIME /24.),2)}\result{} events per day, with \FLARERATIO \% of the data being marked as flaring. The flares had an energy output in the order of $10^{29}-10^{32}$ ergs, originating from at least 
        $\approx 4-30\%$
        of the stellar surface;
    \item A fit to the cumulative flare frequency distribution yields     
        $\alpha=\ALPHALINEAR $, while a maximum likelihood estimator gives $\alpha=\ALPHAME$, in good agreement with previous findings;
    \item Most of the flares were multiple/complex events, two of the events showed 
        quasiperiodic post-flare oscillations with the time scale of a few hours, probably caused by periodic motions of the emitting plasma or oscillatory reconnection;
    \item Superflares (events with energy output over $10^{33}$ ergs) are expected 
        $\approx 3$ times per year, flares a magnitude larger (with $10^{34}$ ergs) every second year -- this could reduce the chances of Proxima Cen b being habitable, as the planet is only 1/20th AU from the host star, and the fluence of radiation and particles increase inversely proportional to the square of the (decreased) distance;
    \item Long-term trends can be seen in both the short-cadence SAP and the 
        long-cadence full-frame images (FFI), that can be compatible with the $\approx 83$ day-long rotation period measured earlier. Unfortunately the length of the dataset does not allow to safely confirm this period;
    \item No obvious signs of planetary transits were detected in the light curve.
\end{itemize}

\acknowledgments
The authors would like to thank the anonymous referee for the helpful comments and suggestions, that significantly improved the paper.
 K.V. is supported by the Bolyai J\'anos Research Scholarship of the Hungarian Academy of Sciences.
 L.v.D.G. is partially funded under STFC consolidated grant No. ST/S000240/1.
The authors thank the financial support  from the National Research, Development and Innovation Office (NKFIH) under grant NKFI-KH 130372.
 This paper includes data collected with the TESS mission, obtained from the MAST data archive at the Space Telescope Science Institute (STScI). Funding for the TESS mission is provided by the NASA Explorer Program. STScI is operated by the Association of Universities for Research in Astronomy, Inc., under NASA contract NAS 5–26555.

%

\vspace{5mm}
\facilities{TESS}

\appendix
\begin{table}[h]
    \caption{Flare parameters for Sector 11 data}
    \centering
    \begin{tabular}{llllll}
\hline\hline
\textnumero{}& Time & Equivalent duration & Energy & Note\\
&[BJD-2457000] &[sec] &[erg]&\\
\hline
1 &  1600.3730 &    3.6778 & 3.647e+30 & \\
2 &  1602.1231 &   11.2875 & 1.119e+31 & \\
3 &  1602.2106 &    3.8165 & 3.784e+30 & \\
4 &  1602.5675 &    1.2674 & 1.257e+30 & \\
5 &  1602.8495 &    1.5051 & 1.492e+30 & \\
6 &  1603.2675 &   20.4766 & 2.030e+31 &double peaked\\
7 &  1603.5759 &    3.3961 & 3.367e+30 & \\
8 &  1603.7037 &    2.3921 & 2.372e+30 & \\
9 &  1603.7926 &    1.0143 & 1.006e+30 & \\
10 &  1604.9051 &    3.3981 & 3.369e+30 & \\
11 &  1605.0009 &    0.8642 & 8.569e+29 & \\
12 &  1605.2648 &    3.7919 & 3.760e+30 &double peaked\\
13 &  1605.6815 &    1.8447 & 1.829e+30 & \\
14 &  1605.9482 &   29.4649 & 2.921e+31 & \\
15 &  1607.4176 &    6.3550 & 6.301e+30 &double peaked\\
16 &  1607.7038 &   13.1791 & 1.307e+31 & \\
17 &  1609.6677 &    2.2111 & 2.192e+30 & \\
18 &  1613.7400 &   87.8975 & 8.715e+31 &triple peaked\\
19 &  1615.4594 &    2.3739 & 2.354e+30 & \\
20 &  1616.0053 &    2.4455 & 2.425e+30 & \\
21 &  1616.0636 &   10.7013 & 1.061e+31 &double peaked\\
22 &  1616.8789 &    4.1078 & 4.073e+30 & \\
23 &  1617.0789 &    5.9920 & 5.941e+30 & \\
24 &  1617.1872 &    9.2374 & 9.159e+30 & \\
25 &  1617.6178 &    2.7855 & 2.762e+30 & \\
26 &  1618.2456 &    2.2871 & 2.268e+30 & \\
27 &  1618.2928 &   27.9062 & 2.767e+31 & \\
28 &  1619.7428 &   35.8418 & 3.554e+31 & \\
29 &  1622.4053 &    9.6765 & 9.594e+30 & \\
30 &  1623.1664 &  175.6592 & 1.742e+32 & \\
31 &  1623.2748 &   14.9430 & 1.482e+31 & \\
\hline
\end{tabular}
\label{tab:energies}
\end{table}

\begin{table}
    \caption{Flare parameters for Sector 12 data}
    \centering
    \begin{tabular}{llllll}
\hline\hline
\textnumero{}& Time & Equivalent duration & Energy & Note\\
&[BJD-2457000] &[sec] &[erg]&\\
\hline
32 &  1625.3539 &   15.4322 & 1.530e+31 & \\
33 &  1625.9150 &    2.0581 & 2.041e+30 & \\
34 &  1626.8428 &   13.0850 & 1.297e+31 & \\
35 &  1629.5886 &    3.8677 & 3.835e+30 & \\
36 &  1630.8136 &    2.6821 & 2.659e+30 & \\
37 &  1630.9220 &    4.3070 & 4.270e+30 & \\
38 &  1632.7275 &   16.0279 & 1.589e+31 & \\
39 &  1633.2622 &   22.9272 & 2.273e+31 & \\
40 &  1633.9580 &    4.1157 & 4.081e+30 & \\
41 &  1636.0441 &    4.1674 & 4.132e+30 & \\
42 &  1636.4108 &   12.4211 & 1.232e+31 & \\
43 &  1637.1274 &    4.3895 & 4.352e+30 & \\
44 &  1637.6927 &   28.4113 & 2.817e+31 & \\
45 &  1640.6927 &    2.1842 & 2.166e+30 & \\
46 &  1640.8704 &    1.5341 & 1.521e+30 & \\
47 &  1641.2218 &   23.0312 & 2.284e+31 & \\
48 &  1641.9954 &    1.0087 & 1.000e+30 & \\
49 &  1642.3704 &    6.7074 & 6.650e+30 & \\
50 &  1643.1204 &    1.3478 & 1.336e+30 & \\
51 &  1643.1884 &    2.7649 & 2.741e+30 & \\
52 &  1643.5037 &    5.2047 & 5.160e+30 & \\
53 &  1644.0134 &    2.9587 & 2.934e+30 & \\
54 &  1644.0398 &    2.9998 & 2.974e+30 & \\
55 &  1644.8842 &    4.2421 & 4.206e+30 & \\
56 &  1645.4092 &    2.4270 & 2.406e+30 & \\
57 &  1645.5301 &   14.5296 & 1.441e+31 & \\
58 &  1647.2078 &    4.6653 & 4.626e+30 & \\
59 &  1647.3008 &   16.5996 & 1.646e+31 & \\
60 &  1647.8119 &    2.0402 & 2.023e+30 & \\
61 &  1648.1605 &   19.4783 & 1.931e+31 & \\
62 &  1648.2494 &   11.0045 & 1.091e+31 & \\
63 &  1649.2619 &    3.6956 & 3.664e+30 & \\
64 &  1649.6494 &    3.2962 & 3.268e+30 & \\
65 &  1650.4008 &    4.4968 & 4.459e+30 & \\
66 &  1650.5563 &    1.9609 & 1.944e+30 & \\
67 &  1651.2327 &   18.4329 & 1.828e+31 & \\
68 &  1651.3702 &    1.4980 & 1.485e+30 & \\
69 &  1651.5188 &    1.8163 & 1.801e+30 & \\
70 &  1651.6368 &    6.4455 & 6.391e+30 & \\
71 &  1652.1938 &   21.7023 & 2.152e+31 & \\
72 &  1652.8799 &    5.3145 & 5.269e+30 & \\
\hline
\end{tabular}
\label{tab:energies2}
\end{table}

\begin{figure}
\plotone{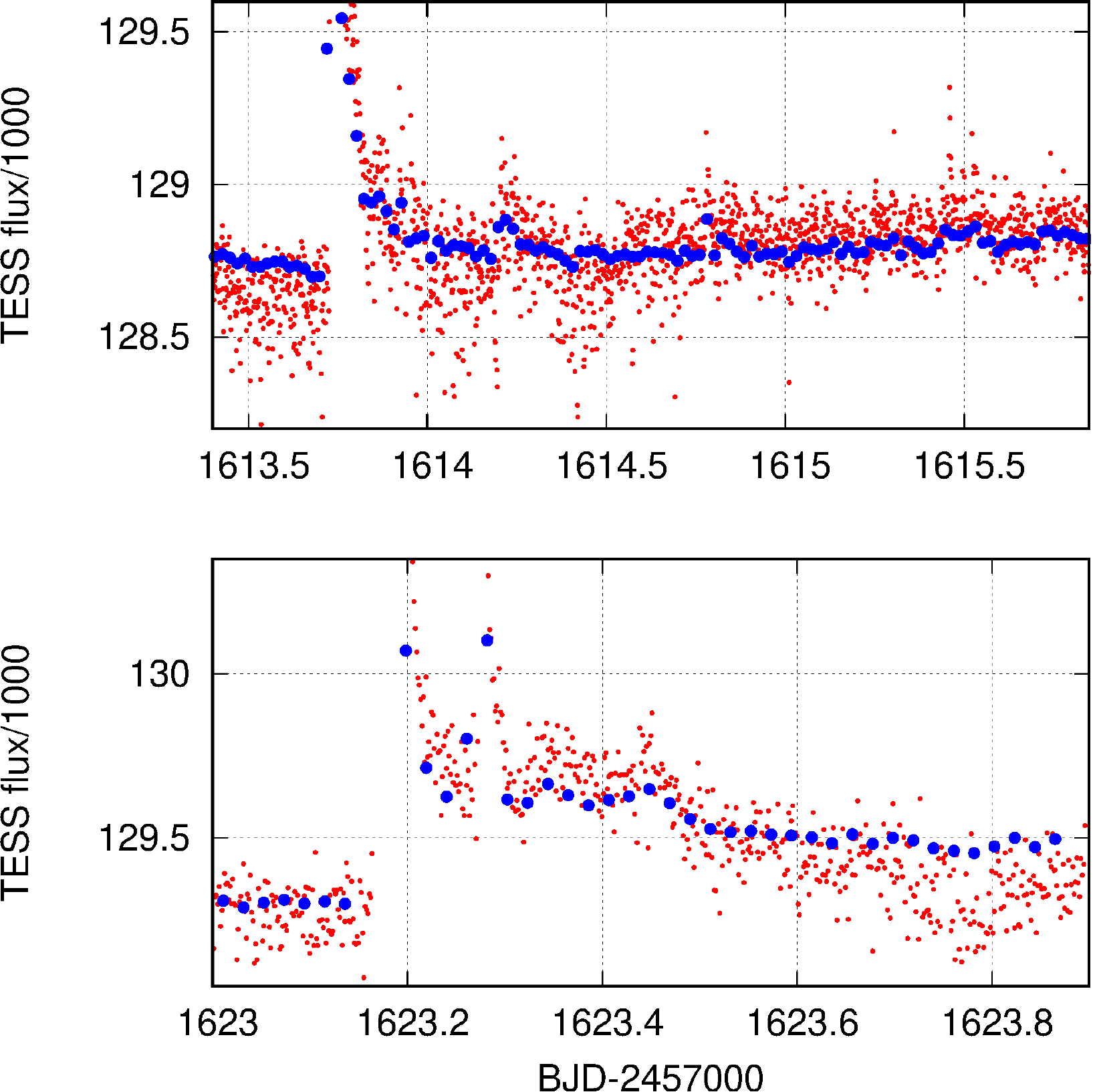}
    \caption{Comparison of the short-cadence simple aperture photometry (SAP) data and long-cadence light curve obtained from photometry of full-frame images. Oscillations on an hourly time scale (see Sect. \ref{sect:oscillations}) are present in both data sets. Note, that not the full range of the light curve is shown. The flux values of the short- and long-cadence data are shifted to match each other. The difference between them is about 3\%, possibly due to the different reduction procedures applied.}
    \label{fig:bumps}
\end{figure}

\begin{figure}
\plotone{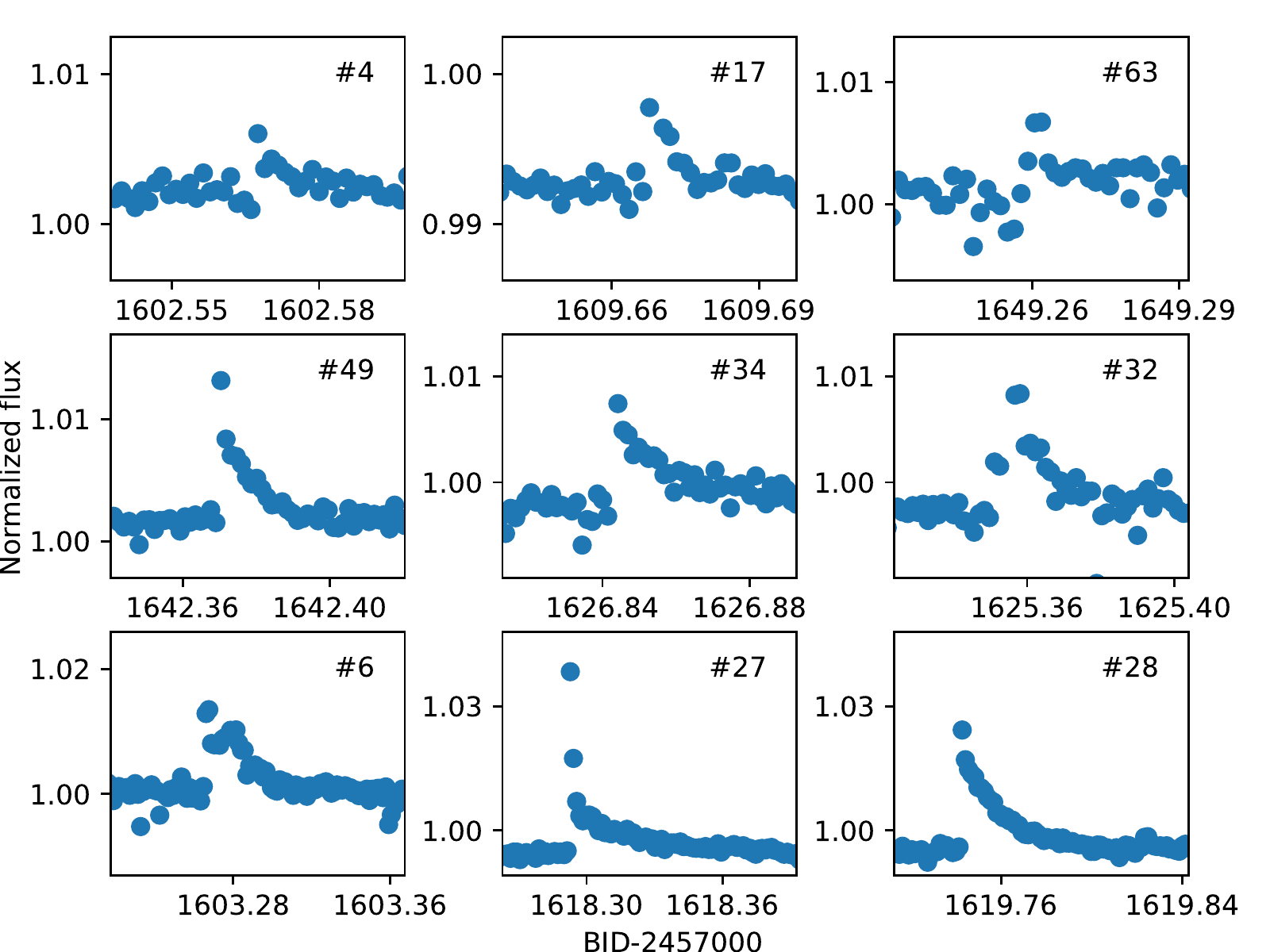}
    \caption{A selection of flares with different energy outputs. Numbers shown in the plots correspond to the event IDs in Table \ref{tab:energies} and \ref{tab:energies2}.}
    \label{fig:flarezoo}
\end{figure}
\end{document}